\documentclass[a4paper]{article}

\usepackage{INTERSPEECH2020}

\title{Static and Dynamic Measures of Active Music Listening as Indicators of Depression Risk}
\name{Aayush Surana$^1$, Yash Goyal$^1$, Vinoo Alluri$^1$}

\address{
  $^1$Cognitive Science Lab, International Institute of Information Technology Hyderabad}
  
\email{\{aayush.surana, yash.goyal\}@research.iiit.ac.in,\\ vinoo.alluri@iiit.ac.in}

\begin{document}

\maketitle
\begin{abstract}
Music, an integral part of our lives, which is not only a source of entertainment but plays an important role in mental well-being by impacting moods, emotions and other affective states. Music preferences and listening strategies have been shown to be associated with the psychological well-being of listeners including internalized symptomatology and depression. However, till date no studies exist that examine time-varying music consumption, in terms of acoustic content, and its association with users’ well-being. In the current study, we aim at unearthing static and dynamic patterns prevalent in active listening behavior of individuals which may be used as indicators of risk for depression. Mental well-being scores and listening histories of 541 Last.fm users were examined. Static and dynamic acoustic and emotion-related features were extracted from each user's listening history and correlated with their mental well-being scores. Results revealed that individuals with greater depression risk resort to higher dependency on music with greater repetitiveness in their listening activity. Furthermore, the affinity of depressed individuals towards music that can be perceived as sad was found to be resistant to change over time. This study has large implications for future work in the area of assessing mental illness risk by exploiting digital footprints of users via online music streaming platforms.
\end{abstract}

\section{Introduction}
Depression has become the leading disability in the modern world affecting around $4.4\%$ of the population worldwide by 2015, according to reports from World Health Organization \cite{world2017depression}. There has been a net increase of $18.4\%$ in the number of people living with depression between 2005 and 2015 \cite{vos2016global} and is only going to increase as a result of the COVID-19 pandemic \cite{sher2020impact, rehman2020depression, ozdin2020levels, mazza2020nationwide}. This has made depression an important topic of research associating it to different behavioral aspects of individuals with major implications on society. Research has demonstrated that during periods of depression, there is a strong reliance on listening to music impacting moods, emotions and altering affect states of an individual \cite{stewart}.

Musical engagement strategies have been linked to measures of ill-health including internalized symptomatology and depression \cite{rumin, litlink}. It has been found that using music for rumination, avoidance and mood worsening is an indicator of risk for depression \cite{hums}. To add to this the inability to cease from repeatedly listening to music which leads to such worsening has also been one of the representative engagement strategies of individuals at risk for depression \cite{hums}.  In terms of musical content, people who are at-risk for depression demonstrate a liking for sad music \cite{garrido}. The vast majority of these studies have relied on self-reported musical consumption data and measures, typically using standard questionnaires, which  may suffer from demand characteristics and/or social desirability bias. Therefore, in this age of big data, examining naturally occurring online consumption, especially using streaming platforms, provides a more ecologically valid reflection of their true preferences and behaviours \cite{nave} and has been proposed as the new frontier \cite{greenberg2017social}. 

Studies have emerged that have examined mental well-being using multimedia content from social networking platforms such as Facebook, Twitter, Instagram \cite{munmun,copper,munmun_14(2),munmun_16,reece}. Till date, Surana et. al \cite{tag2risk} has been the only study that examined association between mental well-being and naturally occurring music listening habits. They examined online music consumption via the music streaming platform Last.fm\footnote{www.last.fm} and associated attributes of the most listened to music tracks with the respective users' well-being scores. Subsequently, the social tags,  specifically adjectives and adverbs, associated with the most frequently listened tracks per user were projected onto the standard 2-dimensional emotion space, that is Valence-Arousal (VA) model or the Russell’s Circumplex Model of Affect [19] where \textit{Valence} represents pleasantness and \textit{Arousal} represents energy or activity. In addition, a 3-dimensional emotion model (VAD) was also used with the addition of the \textit{Dominance} or perceived intensity dimension and compared results. They found that individuals at-risk for depression predominantly listen to music which are socially tagged as being representative of low arousal and negatively valenced emotion, sadness, and its related terms, such as \textit{dead}, \textit{low}, \textit{depressed}, \textit{repetitive}, and \textit{miserable}. 
However, solely focusing on social tags does not yield insights into the acoustic properties of the music. For example, a heavy metal piece of music with tags labelled as \textit{transcendental} or \textit{wonderful} and a sad love song or ballad with the same tags are different acoustically but would get clubbed into the same emotion category. Furthermore, a considerable amount of tracks on online platforms remain untagged resulting in loss of information. Hence it is important to examine acoustic features of music and the perceived emotions extracted from them thereof to get a better understanding of music consumption of individuals at-risk for depression. 

Furthermore, emotions are dynamic processes that fluctuate in response to environmental demands and regulatory forces \cite{koval2013affect} and certain patterns or irregularities with which they fluctuate across time have been associated with psychological well-being. Higher levels of emotion variability and inertia have been linked to depression \cite{kuppens2017emotion, houben2015relation}. These fluctuations in the internal emotion states of individuals might be reflected in their dynamic use of music. Hence, it is important to investigate time-varying musical consumption.

\begin{figure*}[!ht]
  \centering
  \includegraphics[width=0.9\textwidth]{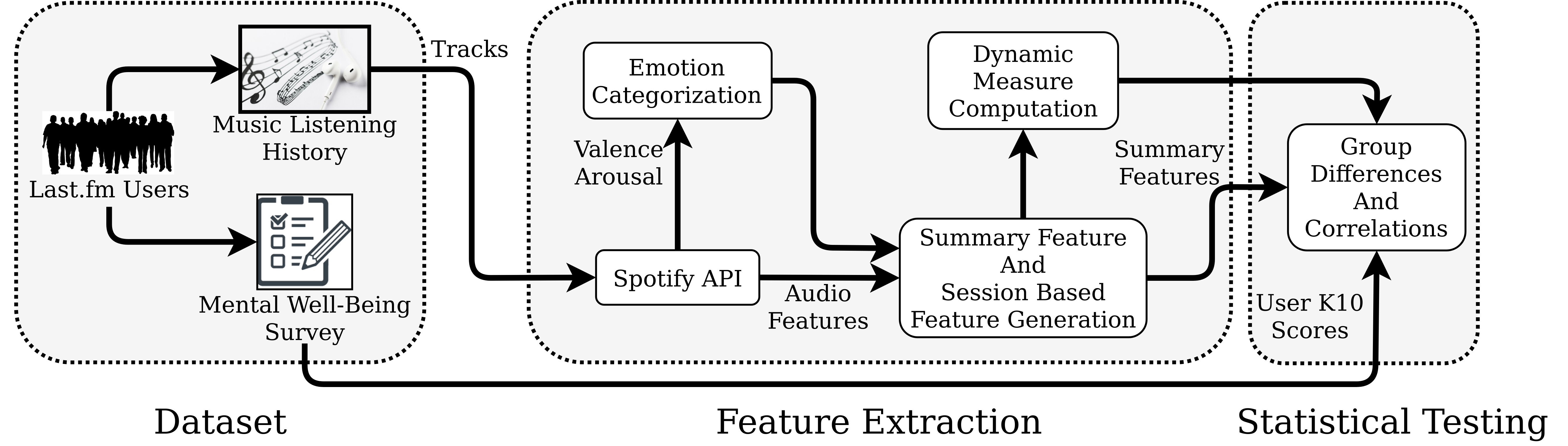}
  \caption{Methodology}
  \label{methodology}
\end{figure*}

The aim of this study is to characterize online music consumption of individuals at-risk for depression by analyzing acoustic and emotion-related features and their dynamic variation.
To this end, we analyze naturally occurring music listening behavior via listening histories of Last.fm users. We also examine time-varying properties of acoustic and emotion-related features and associate them with the respective users' well-being scores indicative of risk for depression. Based on previous literature we hypothesize the following Last.fm users who demonstrate high depression risk that is characterized by low well-being scores:
\begin{itemize}
 \item greater overall consumption of music, especially music that is perceived as sad
  \item higher repetition in music tracks owing to ruminative and repetitive engagement. 
   \item increased level of variability and/or inertia in terms of the acoustic and emotion information of music across time 
\end{itemize}
The results are discussed in the context of potential underlying moods or states.

\section{Methodology}
Figure 1 encapsulates the procedure used in our study and is described as follows.

\subsection{Dataset}
Data was from the aforementioned Surana et al. \cite{tag2risk} study comprising listening histories of 541 Last.fm users (82 females, mean age = 25.4 years, std = 7.3 years). The respective users' well-being was measured using standard diagnostic questionnaire, the Kessler’s Psychological Distress Scale (K10) \cite{k10}. Additionally, the Healthy-Unhealthy music (HUMS) \cite{hums} scale was administered to measure musical engagement which describes music consumption behavior and indirectly indicates risk for depression, in addition to personality data. 
HUMS and personality information have not been used in the current study, but have been additionally used in the original study \cite{tag2risk} to evaluate internal consistency, which was found to be high. K10 scores demonstrated high reliability (cronbach alpha $= 0.91$). The participants' listening histories were extracted for a duration of 6 months centered (± 3 months) around the time they filled the K10 questionnaire. 




\subsection{Feature Extraction}
\begin{figure}[h]
 \centerline{
 \includegraphics[width=0.75\columnwidth]{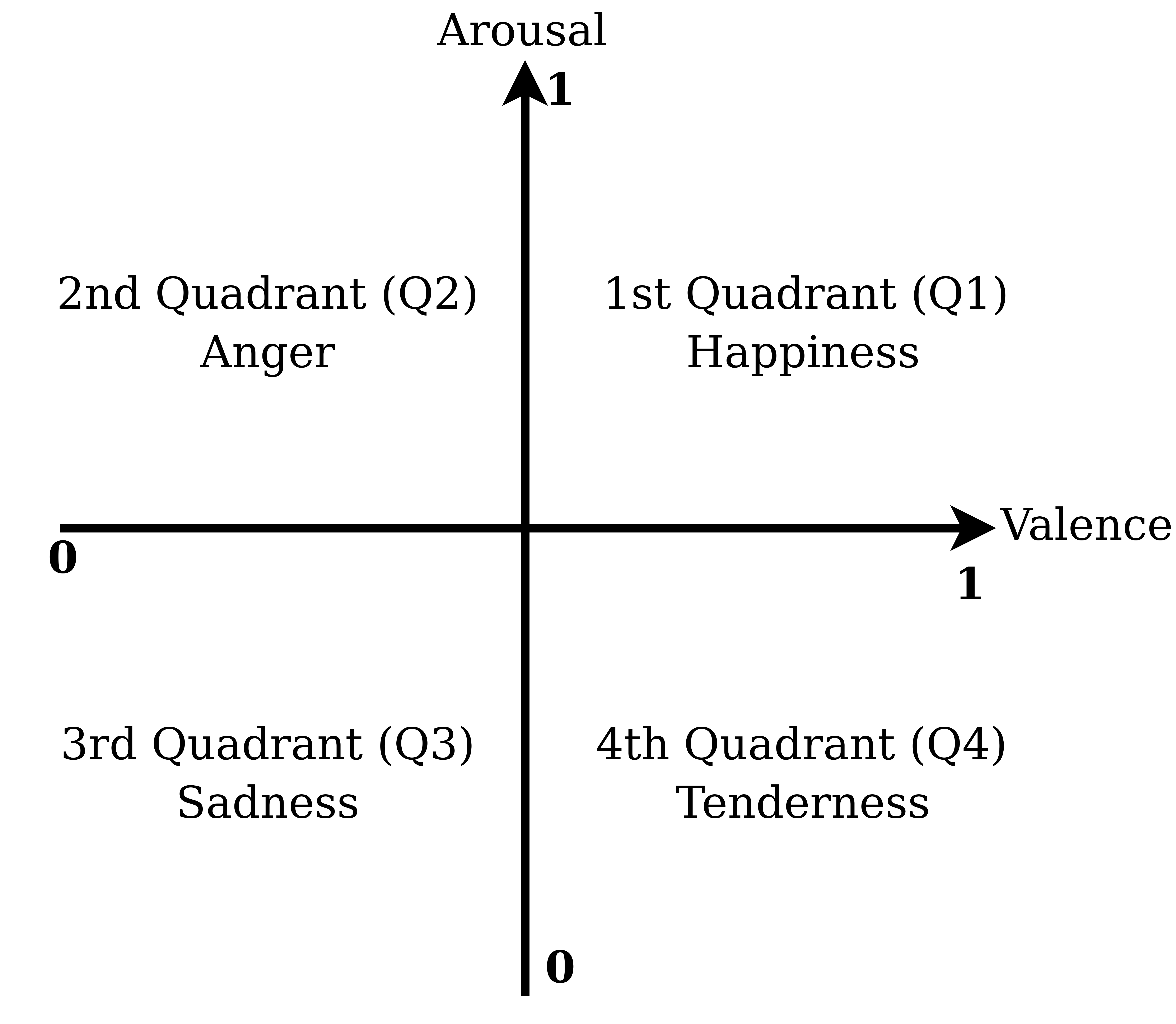}}
 \caption{Four quadrants of Valence-Arousal space representing four emotional states.}
 \label{VAspace}
\end{figure}

In order to extract acoustic features from a track we used the Spotify\footnote{www.spotify.com} public API. Spotipy\footnote{developer.spotify.com} package was employed to search for each track in Spotify database and retrieve the values for 10 features that Spotipy provides. Eight of these were audio features which comprised \textit{danceability, loudness, speechiness, acousticness, instrumentalness, liveness, tempo}, and \textit{mode}. All the features take continuous values except mode which is binary (0:major, 1:minor). Additional two emotion features representing the \textit{valence} and \textit{energy/arousal} were also retrieved. We further categorized each track into one of the four quadrants of the 2-dimensional VA space \cite{VAmodel}. The first quadrant loosely translates to the emotion category representing \textit{happiness}, the second quadrant to \textit{anger}, the third to \textit{sadness} and the fourth to \textit{tenderness}, as shown in Figure \ref{VAspace}\cite{VAmodel}. For each quadrant we calculated a quadrant prevalance score (QPS) which is calculated as the proportion of tracks in the respective quadrants in each user's listening history. As mentioned before, individuals at risk for depression have an inclination for music representing sadness \cite{tag2risk}. Hence, in order to get a bird's eye-view of the acoustic features associated with individuals at-risk when compared to no risk for depression, we calculated static features that comprise mean score for the eight audio features as well as the QPS over each user’s entire listening history which we term as \textit{summary features} hereon.

\subsubsection{Session-based feature generation}
Since we aimed to capture the changes in participants' audio features and emotions over time, we defined time resolution based on sessions. A session was defined as a period of continuous listening activity, with a subsequent session occurring after an inactivity of at least two hours \cite{gupta}. Further, in each session, audio features for all the tracks were averaged to denote the value of that feature for the respective session. Also, the QPS for the four quadrants were computed per session for each user. Thus, we obtained time-varying values for both, audio and QPS features, for each user over his/her 6-month listening period. The reason for using a smaller time interval/resolution based on sessions as opposed to larger intervals like days or weeks is due to the fact that emotions are more transient and short-lived than mood states which may span for days \cite{rottenberg2005mood}. For example, continued music consumption across sessions with certain features may result in greater successive session-based QPS in certain quadrants and might reflect underlying moods. For example, a user might continuously listen to sad music, the acoustic features of which might be predominant in say the third quadrant representing sadness, and hence may be indicative of underlying risk for depression based on aforementioned previous literature. Furthermore, we extracted additional measures that capture such dynamic behavior which provide a better understanding of users' mood shifts characterized by variability and inertia. These additional dynamic measures are explained in following section.

Finally, since we had hypothesized that individuals at-risk for depression may engage in repetitive music listening behaviors, we defined a \textit{repetitiveness index} (RI) per user, computed as shown in Equation \ref{eq1}. In order to capture a user's repetitiveness within a session, the proportion of tracks that were repeated is first calculated. Subsequently, RI is calculated as the average \textit{repetitiveness} across all sessions.
\begin{equation}\label{eq1}
RI_{u}=\frac{1}{N_{u}}\sum_{j\epsilon{N_{u}}} \frac{\sum_{i\epsilon{T_{u,j}}}\left(tr\textsubscript{i,j}\times I\left(tr\textsubscript{i,j}>=2\right)\right)}{\sum_{i\epsilon{T_{u,j}}}tr\textsubscript{i,j}}
\end{equation}
where,\\
$N_{u}$ : number of sessions for user $u$\\
$T_{u,j}$ : all unique tracks in $jth$ session for user $u$\\
$tr_{i,j}$ : frequency of track $i$ in $jth$ session \\
$I(condition)$ : indicator function, 1 if $condition$ evaluates to true, else 0\\
\subsubsection{Dynamic Measures Computation}
Previous studies have demonstrated the association of dynamic measures such as variability, instability, and inertia, with depression\cite{seabrook2018predicting}. Variability represents the standard deviation of scores across sessions. Inertia is calculated as the autocorrelation coefficient (with time lag of 1 session) for the time-varying scores. It captures the predictability of a feature/emotion from one time-point to the next. Higher inertia and extreme variability have been found to be representative of ill-being and depressive moods, especially in negative affect states \cite{kuppens2017emotion}. Instability is calculated as the mean squared successive difference (MSSD) for scores across sessions. It measures the deviation in scores from one time-point to the next. A person with high instability implies higher fluctuations in music consumption with respect to the features/emotions. Since, instability has been found to have mathematical dependency on the other two measures, that is, it is directly proportional to variability and has inverse relation with inertia \cite{koval2013affect}, we restrict our analyses to computing only variability and inertia. These two dynamic measures were calculated for all the session-based acoustic features and QPS.

\subsection{Statistical Testing}
We first performed tests of difference by dividing the users into No-Risk and At-Risk groups based on their K10 scores as done in Surana et al. \cite{tag2risk}. Users with K10 $< 20$ (n=193) fall under the No-Risk group while K10 $> 29$ (n=142) comprised the At-Risk group. We performed a two-tailed Mann Whitney U (MWU) test on the static \textit{summary features} between these groups. 
We further performed bootstrapping to account for Type I error and ensure that the observed differences are not due to chance using the same approach as Surana et al. 
Next we performed Spearman correlations between K10 and all 541 users' number of sessions, total playcount\footnote{Here, the total playcount refers to the total number of tracks(including the repetitions) listened by the user.}, and \textit{RI}. Similarly, correlation between K10 and each of the dynamic measures of audio and emotion features was computed. We additionally performed permutation testing to account for multiple comparisons due to the number of correlations performed and hence account for Type 1 errors. For each correlation, permutation testing was performed with 10,000 iterations. At each iteration, the K10 scores were randomly assigned (with replacement) to each user. The correlation value for each iteration was calculated. The significance of the original statistic (Spearman correlation) was estimated from the distribution of the correlation statistics thus obtained.

\section{Results}
\begin{figure}[h]
 \centerline{
 \includegraphics[width=0.8\columnwidth]{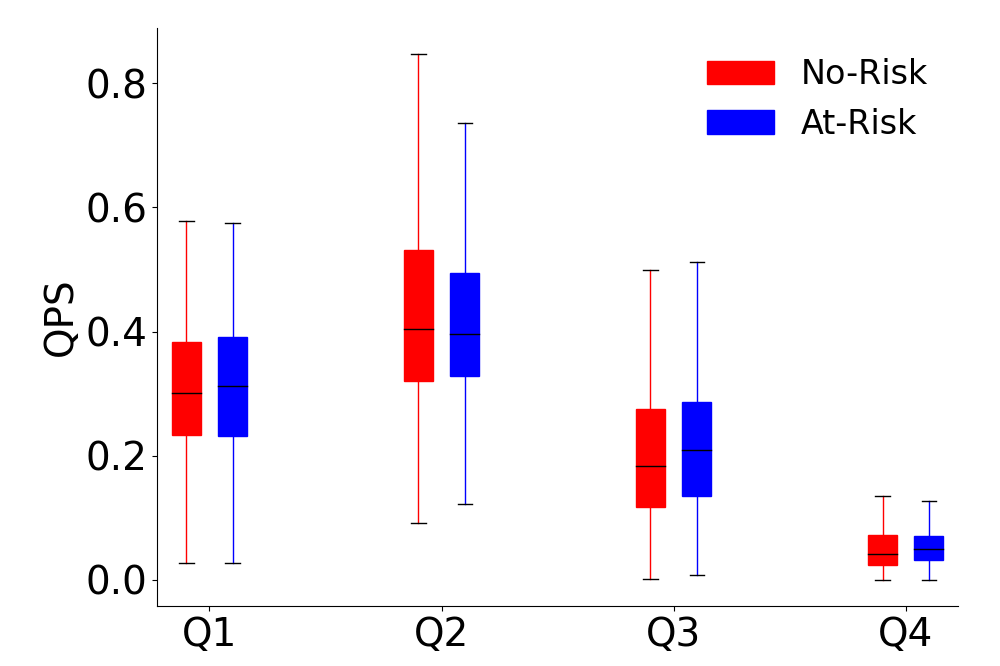}}
 \caption{Boxplot of QPS \textit{summary feature} (for each of the four quadrants) for No-Risk and At-Risk groups.}
 \label{boxplot}
\end{figure}

There were about 8,09,000 unique tracks in the dataset, out of which the audio features were available for 5,73,000 (~71\%) tracks on Spotify. No significant differences were observed between the No-Risk and At-Risk groups using MWU test based on the static \textit{summary features}. A borderline significant difference ($p=0.06$) was observed for the acoustic feature \textit{speechiness} having a higher median for the At-Risk group. Figure \ref{boxplot} displays the boxplots of the QPS \textit{summary feature} for both groups.
A borderline group difference ($p=0.1$) for the QPS in the fourth quadrant (Q4) representing \textit{Tenderness}. However, we do observe that the median in the third quadrant (Q3) representing \textit{Sadness} was higher for the At-Risk group.

On an average, each user had 292 sessions. 
Significant positive correlations were observed between K10 and number of sessions ($r=0.16,p=0.0003$), total playcount ($r=0.11,p=0.01$) and \textit{RI} ($r=0.16,p=0.0002$). These correlations were found to be significant post permutation testing (all $p < .01$).
\begin{table}[h]
\centering
\begin{tabular}{|c|c|c|c|c|}
\hline
Audio Feature&Variability&Inertia\\
\hline
Danceability&-0.03&0.03\\
Loudness&-0.01&0.05\\
Speechiness&0.05&0.09*\\
Acousticness&0.02&0.05\\
Instrumentalness&-0.11**&0.02\\
Liveness&-0.05&0.05\\
Tempo&-0.05&0.03\\
Mode&-0.05&0.07\\
\hline
\end{tabular}
\caption{Spearman correlations between users' dynamic measures represented by Variability and Inertia of audio features and K10 scores. (*p\textless0.05, **p\textless0.01)}
\label{tab:audio}
\end{table}

Table \ref{tab:audio} displays the correlation results between the dynamic measures of the \textit{audio features} and K10 scores. K10 had a negative correlation with variability in \textit{instrumentalness} ($r=-0.11,p=0.009$) which retained significance even after controlling for mean and inertia. Also, K10 correlated positively with inertia in \textit{speechiness} ($r=0.09,p=0.03$) retaining significance after controlling for mean and variability. These correlations were also found to be significant after permutation testing.

\begin{table}[h]
\centering
\begin{tabular}{|c|c|c|c|c|}
\hline
Quadrant&Variability&Inertia\\
\hline
Q1&-0.06&0.03\\
Q2&-0.06&-0.03\\
Q3&0.02&0.1*\\
Q4&0.03&0.05\\
\hline
\end{tabular}
\caption{Spearman correlations between users' dynamic measures, represented by Variability and Inertia of QPS and K10 scores. (*p\textless0.05)}
\label{tab:quadrant}
\end{table}
Table \ref{tab:quadrant} presents the correlation results between the dynamic measures of QPS and K10. As expected, significant positive correlation was observed between inertia and K10 in the third quadrant (Q3) representing \textit{sadness} ($p=0.02$) retaining significance post permutation testing. In addition, this correlation was preserved after controlling for mean QPS ($r=0.09,p=0.03$).

\section{Discussion}
This study is the first to analyze the association between depression risk and music consumption via acoustic features. We analyse naturally occurring listening behavior of individuals on the online music streaming platform Last.fm as opposed to self-reported or lab-based studies. Further, this study is the first of its kind to look at the dynamic nature of music consumption to characterize risk for depression.

No significant differences were found between At-Risk and No-Risk groups in terms of audio features when they were averaged over the entire listening history of participants. This suggests that the kinds of music listened by individuals with varying depression risk has comparable features. However, a further inquiry into genre preferences might shed light on perceptual attributes that might be more relevant and representative of At-Risk individuals' consumption. The borderline significant difference demonstrated in the fourth quadrant QPS and sub-threshold observable difference between the median QPS and respective ranges in Q3 (Figure \ref{boxplot}) are very much in line with previous studies \cite{garrido, tag2risk}. Individuals with high K10 scores are characterized by states that are high in anxiety and arousal. When using music as a coping mechanism, a natural choice may be to listen to music that is low on arousal owing to their already heightened physiologically aroused states. Since, we take a subset of the participants to investigate group differences, future studies can aim to increase sample size to unearth more definitive trends.

Session-based analyses yielded several significant results which are in line with previous studies. The positive correlation observed between K10 and the total number of sessions and playcount suggests greater reliance on music for individuals with higher risk for depression. Higher musical engagement is representative of emotion focused-coping \cite{miranda2012music}, especially during periods of depression \cite{stewart}.
Furthermore, the positive association between \textit{Repetitiveness Index} and K10 reflects the ruminative coping style which is characteristic of individuals at-risk for depression \cite{hums} which may result in intensifying their negative moods \cite{garrido, garrido2015moody}. Whether this ruminatory behaviour with music as a virtual other is beneficial or maladaptive in the long-run remains to be seen.

The negative association found between K10 and variability in \textit{instrumentalness}, an acoustic feature that indicates the instrumental nature or lower presence of spoken words in songs, suggests that the greater the risk of depression, the less the deviation in \textit{instrumentalness}
This result coupled with higher median \textit{instrumentalness} for At-Risk group than No-Risk is in line with the findings of Surana et al. \cite{tag2risk} who report that At-Risk individuals gravitated towards neo-psychedelic-, avant garde-, dream-pop genres that are characterized by obscure vocals. The positive association between Inertia in \textit{speechiness}, an acoustic feature that captures presence of spoken words in a track, and depression risk suggests that high risk individuals tend to be resistant to change when listening to certain music associated with specific levels of \textit{speechiness} over time. One limitation is the of assignment an average estimate of an acoustic feature to a track, since it does not capture within-track variability. Hence, two tracks which sound perceptually different may have similar values for an acoustic feature as a result of averaging over the track. However, combining acoustic features with tag information in future may provide relevant perceptual information and hence better insight into describing aspects of music associated with individuals at risk.

Finally, the positive association between inertia in the Q3, representing \textit{Sadness} in the VA space, and K10 supports our hypothesis that individuals At-Risk not only listen to music that is more reflective of their negatively valenced states, but also that this affinity is more resistant to change over time. Since it is known that individuals at risk for depression exhibit higher inertia in their own internal states, we can conclude that this behavior is reflected in their music consumption patterns. This is a novel finding that further corroborates the notion that music is indeed a mirror of the self, especially in terms of dynamic states.

 
In future, we intend to employ a multimodal approach of incorporating tags, acoustic features, and lyrics-based information such as sentiments, to gain a holistic understanding of music consumption characterizing depression risk. This would then pave way to designing early depression detection systems using digital music footprints.

\bibliographystyle{IEEEtran}

\bibliography{mybib}


\end{document}